\input{aipcheck}

\documentclass[
    ,final            
  ]
  {aipproc}

\layoutstyle{6x9}

\def\be{\begin{equation}}

\def\ee{\end{equation}}

\def\bea{\begin{eqnarray}}

\def\eea{\end{eqnarray}}

\def\bi{\begin{itemize}}

\def\ei{\end{itemize}}

\begin{document}

\title{Accurate determination of the superfluid-insulator transition in the one-dimensional Bose-Hubbard model}

\classification{03.75.Lm, 73.43.Nq}
\keywords      {quantum phase transition, numerical methods}

\author{Jakub Zakrzewski}{
  address={Instytut Fizyki imienia Mariana Smoluchowskiego and Mark Kac Complex
Systems Research Center, Uniwersytet Jagiello\'nski, ulica Reymonta 4,
PL-30-059 Krak\'ow, Poland}
}

\author{Dominique Delande}{
  address={Laboratoire Kastler-Brossel, Universit\'e Pierre et Marie Curie-Paris 6, ENS, CNRS;  4 Place Jussieu, F-75005 Paris, France}
}

\begin{abstract}
The quantum phase transition point between the insulator and the superfluid phase 
at unit filling factor of the infinite one-dimensional Bose-Hubbard model is numerically
computed with a high accuracy. The method uses the infinite system version of the time evolving block decimation algorithm, 
here tested in a challenging case. We provide also the accurate estimate of the phase transition point at double occupancy.
\end{abstract}

\maketitle


\section{Introduction}

Ultra-cold atoms in optical lattices make possible detailed studies of basic condensed
matter systems \cite{jaksch,bloch,esslinger,annals}.
While several applications have been envisaged ranging from 
 high $T_c$ superconductivity \cite{hightc},
disordered systems \cite{sanpera}, spin models \cite{duan} or 
novel quantum magnets \cite{bodzio1} - for a review, see \cite{lewen1} -
the most basic model realized in such a setting is the Bose-Hubbard model (BHM) \cite{fisher},
which still defies an exact analytical
description while exhibiting a nontrivial
quantum phase transition \cite{sachdev} between a superfluid state and an insulator.
Experiments, both three-dimensional \cite{bloch} and
one-dimensional (1D) \cite{esslinger}, performed in additional confining atomic traps, 
led to realization of an inhomogeneous finite model.
Still, from the theoretical perspective, the infinite
system is the most challenging case.
As in experiments, the typical numerical studies deal with finite size systems, with
a possible extrapolation to infinite systems at the end. Such an extrapolation is always
delicate, especially in the vicinity of the phase transition. Even simple quantities like
the precise location of the phase transition point are still controversial, see below. 

In the absence of an exact analytical result, one must rely on approximate results,
such as the Bethe ansatz~\cite{krauth}
or numerical studies, which are roughly of three different types: exact
diagonalization for finite systems - limited to very small systems and thus unable
to extrapolate to the infinite size -, 
Quantum Monte-Carlo (QMC) methods and Density Matrix Renormalization Group (DMRG) methods \cite{white}. 

The Time Evolving Block Decimation (TEBD) algorithm \cite{vidal1,vidal2} is a novel variant
of the DMRG method which can provide results for the temporal dynamics, but also
for the eigenstates using imaginary time evolution.
Recently, it has been extended to
1D lattice systems with nearest-neighbor interactions \emph{of infinite size}~\cite{vidali,orus,mcculloch}.  
It has been successfully tested on the integrable
Ising model. It is yet to be seen whether really difficult problems can be attacked by this algorithm. 
The present work provides a partial answer to this question
by addressing the challenging problem of the quantum phase transition of the BHM.

\section{The Bose-Hubbard phase transition}
The Hamiltonian of the BHM, in dimensionless variables, reads
\begin{equation}
\label{H}
H= -J \sum_{r=1}^M(\hat{a}_{r+1}^\dag \hat{a}_r + {\rm h. c.})
+\frac{U}{2} \sum_{r=1}^M\hat{n}_r(\hat{n}_r-1) - \mu \sum_{r=1}^M\hat{n}_r,
\end{equation}
where  $M$ is the number of sites. We put it finite just for the introduction, as we shall consider
$M=\infty$ for the actual calculation. $\hat{a}_{r}^\dag$ ($\hat{a}_r $) is the creation (annihilation)
operator at site $r$, while $\hat{n}_r=\hat{a}_{r}^\dag\hat{a}_r$. The total number of atoms
$\hat{N}=\sum_{r=1}^M\hat{n}_r$ commutes with the Hamiltonian. In the thermodynamic
limit of infinite systems, it is convenient to consider the particle density 
$\rho=\langle \hat{N}/M \rangle$ and the chemical potential $\mu.$

Two parameters describe the model: the tunnelling rate $J$ and the interaction
strength $U$. 
Since physics of the BHM depends on the $J/U$ ratio only, we set
$U\equiv1$ for convenience. 
The behaviour of the system as $J/U$ is varied has been extensively discussed in a number of papers,
see~\cite{fisher,krauth,batrouni,singh92,kks,ks,eko,elstner,pai,km,kwm00,pollet,my}, thus we 
mention the most relevant details only.
The Mott insulator phase occurs at integer occupation of sites for sufficiently small tunnelling. 
Within the Mott insulator regime, the density stays constant - the system is not compressible.
Typically, at the phase boundary, the density changes and the superfluid regime is reached via
a standard quantum phase transition. 
The interesting situation occurs when the phase transition is
reached via a path of a constant integer density $\rho$.
The model belongs then to the universality class of
the $XY$ model \cite{fisher} and the phase transition is of the Kosterlitz-Thouless (KT) 
\cite{kosterlitz,haldane} type. 

In the present paper, we determine very accurately the  
 $J$ value at which the KT transition occurs for $\rho=1.$ This is a difficult
task previously addressed using different methods yielding
quite different results. 
The mean field theory \cite{sachdev,zwerger} yields
 $J_c\approx 0.086$. The Bethe ansatz allows for an analytical albeit approximate 
 treatment\cite{krauth} giving  $J_c=1/(2\sqrt{3})\approx
 0.289$. Early numerical values obtained via QMC \cite{batrouni}
 as well as renormalization group arguments \cite{singh92} yielded a quite
 different value $J_c=0.215$.  Later numerical approaches came closer to the
 Bethe ansatz result. In particular, Kashurnikov et al.~\cite {kks} found  $J_c=0.300 \pm 0.005$ in a QMC study
 while an exact diagonalization for $M=12$ combined with renormalization group by the same group \cite{ks}
 yields $J_c=0.304 \pm 0.002$. Other QMC approach \cite{pollet} gave $J_c=0.305\pm0.004$. Earlier diagonalization results~\cite{eko} gave $J_c\approx 0.275 \pm 0.005$. 
 High order perturbative expansion suggests
$J_c=0.26 \pm 0.01$ \cite{elstner} in an excellent agreement with finite DMRG treatment 
\cite{rapsch99} yielding essentially the same result. Yet other DMRG approaches gave
interestingly different results.
The periodic boundary conditions DMRG  \cite{pai} gave $J_c\approx 0.298$ while the infinite
size DMRG with periodic boundaries gave $J_c=0.277 \pm 0.01$ \cite{km}. The same authors attempted
a finite size DMRG with open boundary conditions (yielding a better convergence) giving $J_c=0.297
\pm 0.01$ \cite{kwm00}. 
Therefore, the current algorithms disagree on the
the value of $J_c.$ This is partially due to logarithmic finite-size effects close to
the critical point of the KT transition \cite{ks,kwm00}. The accurate determination
of $J_c$ is thus an ideal testbed for algorithms dealing with infinite systems.

\section{Time-Evolving Block Decimation Algorithm}
The approach \cite{vidali} used is
an infinite variant of the TEBD algorithm~\cite{vidal1,vidal2}.
It belongs to a rapidly developing class of methods which utilise 
the understanding of quantum entanglement in complex systems to facilitate effective
simulations of many-body quantum systems. All these methods have some links 
(see discussion in \cite{daley}) with the DMRG
algorithms \cite{white} extending them either to time evolution \cite{vidal1,vidal2} 
and to many dimensional
systems \cite{cirac,cincio08,jordan})  
A particularly interesting variant of the TEBD approach assumes
translational invariance of the wavefunction and enables a treatment of explicitly
infinite systems \cite{vidali,orus,mcculloch}.
While we refer the reader to original papers for details, we
sketch here the main idea. 

The BHM Hamiltonian involves only
nearest-neighbor interactions and can thus be written as a sum over consecutive sites $r$ of
reduced hamiltonians $h^{(r,r+1)}.$  
Each site is described by a Hilbert space of dimension $D.$ 
We follow the imaginary time evolution, 
i.e., compute (we assume $\hbar=1$ in the whole paper)
\be |\Psi(\tau)\rangle=\frac{\exp(-H\tau)|\Psi(0)\rangle}{||\exp(-H\tau)|\Psi(0)\rangle||}.
\label{psitau}
\ee
For large $\tau$ and a generic initial guess $|\Psi(0)\rangle,$ such a procedure should 
yield a good approximation of the ground state provided there is a gap in the spectrum of the system.

 We assume translational
invariance of our system, i.e., $ h^{(r,r+1)}$'s does not depend on site $r$. Moreover we
restrict ourselves to wavefunctions that are invariant under shifts by
one lattice spacing\footnote{This assumption is not essential for the method and 
can be easily lifted, as pointed out in \cite{vidali}. 
It excludes ``charge density wave'' or ``checkerboard''
solutions. Previous studies of the BHM indicate that the ground state is
translationally invariant so there is no real limitation in the approach used.}. 
The time evolution proceeds in the following way. 
The normalized
wave function is Schmidt decomposed at site $r$ as
\be 
|\Psi\rangle=\sum_{\alpha=1}^{\chi}\lambda_\alpha^{(r)}
|\Phi_\alpha^{(\triangleleft r)}\rangle \otimes  
|\Phi_\alpha^{(\triangleright r+1)}\rangle;
\ee
where $\chi$ is a finite Schmidt rank, the $|\Phi_{\alpha}\rangle$ states are an orthonormal
set and $\sum_\alpha (\lambda_\alpha^r)^2=1$ for global normalization. 
As discussed in \cite{vidal1,vidal2}, for low lying states, the $\lambda_\alpha$ 
values decrease rapidly, so that a good approximation
of the state is obtained keeping only the few largest $\lambda_\alpha^r$. 
The two consecutive left or right
sub lattices states are related via tensors $\Gamma^{(r)}$ as follows:
\bea
|\Phi_\alpha^{(\triangleleft r+1)}\rangle &=& \sum_{\beta=1}^{\chi} 
\sum_{i=1}^{D}\lambda_\beta^{(r)}\Gamma_{i\beta\alpha}^{(r+1)}
|\Phi_\beta^{(\triangleleft r)}\rangle |i^{(r+1)}\rangle,\cr
|\Phi_\alpha^{(\triangleright r)}\rangle &=& \sum_{\beta=1}^{\chi} 
\sum_{i=1}^{D}\Gamma_{i\alpha\beta}^{(r+1)}\lambda_\beta^{(r+1)}
 |i^{(r)}\rangle|\Phi_\beta^{(\triangleright r+1)}\rangle;
 \label{expansion}
   \eea
where $|i^{(r)}\rangle$ is an orthonormal basis spanning the local Hilbert space at site $r$ (e.g. the Fock basis
 for the BHM). In that way
 one obtains a so-called matrix product state (MPS) \cite{fannes}
 representation of $|\Psi(\tau)\rangle$. The imaginary time evolution operation
 can be expressed as a product of elementary two-site evolution operators
 \be U^{(r,r+1)} = \exp{(-h^{(r,r+1)}\delta t}), \qquad \delta t \ll 1;
 \ee
 using the Suzuki-Trotter decomposition \cite{suzuki}. 

Now the crucial observation is that, for translationally invariant $|\Psi\rangle$,
all $\Gamma^{(r)}$ and $\lambda^{(r)}$ are independent of $r$. The application
 of a two-site gate  $U^{(r,r+1)}$ has two effects:
 \bi
 \item it breaks for a moment the translational invariance (since sites $r$ and $r+1$ are updated only);
 \item the resulting vector is not longer a MPS.
 \ei
  The latter is taken care of by
 a singular value decomposition of the resulting matrix and  retaining only the 
$\chi$ largest eigenvalues (and corresponding vectors) - see \cite{vidal1,vidal2} for details
 and \cite{vidali} for a graphical representation. This approximation brings the vector back into the MPS form. 
 The modified $\Gamma^{(r)}$'s, $\lambda^{(r)}$'s
 are then updated for all the sites -- the translational invariance is restored.
  
It is worth stressing that the repetition of this procedure allows one to act on two
consecutive sites only, 
taking proper care of local updates of appropriate tensors. In the limit $\chi \to \infty$ and
$\delta t \to 0,$ the exact evolution is obtained. The error introduced by finiteness of $\delta t$
is easily controlled (it varies as a power of $\delta t$ depending on the
order of the  Suzuki-Trotter decomposition) while the strength of the method is that
the number $\chi$ of not too small $\lambda$ values remains reasonably small. As shown in \cite{vidali}, 
to facilitate convergence, one can take successively smaller and smaller time 
steps. In the following we use a simple second order Trotter expansion (which, according to
our experience,  seems to be the most efficient in imaginary time evolution when time steps are
decreased).

\section{Aplication to the Bose-Hubbard model}
To make the infinite-TEBD method work,
 additional parameters have to be adjusted. One of them is the maximal allowed 
 occupation $N_{max}$ at a given site (giving the dimension of the Hilbert 
 space at each site, $D=N_{max}+1$).  In earlier DMRG calculations for mean unit 
 occupancy typically $N_{max}=4$ or $N_{max}=6$ was used \cite{rapsch99,kwm00}. We take
$N_{max}=6$, although the difference with $N_{max}=4$ is small.
The second important cutoff parameter is the restriction on $\chi$, 
the number of $\lambda$
values retained at each step. This value can be quite small deep in the 
Mott insulator region, it has to be taken  painfully large close to the 
KT transitions. This is costly since the
execution time scales as $\chi^3$~\cite{vidali}. A satisfactory convergence 
of the correlation function
\be
C(s) = <\hat{a}_{s+r}^\dag \hat{a}_r> 
\ee
requires $\chi$ of the order of one hundred.  Such correlation functions 
define the phase coherence of the condensate, we shall use
them to determine the transition point.

In the Mott regime, the system has a finite gap, which implies that $|\Psi(\tau)\rangle$,
eq.~(\ref{psitau}),
converges exponentially fast at long times to the ground state,
at a rate proportional to the energy gap. This is indeed what we numerically
observe, with exponential convergence of the energy too.
Thus, gor a given time step 
we perform the imaginary time evolution until the slope of the energy curve w.r.t. imaginary time 
reaches a desired small threshold. Then the time step is decreased by a factor 2 and
the procedure repeated until a computer accuracy limit for the time step is reached. While
this procedure may not be optimal for speed, it assures the convergence of the results
obtained.
  
\begin{figure}[t]
 \includegraphics[height=.5\textheight]{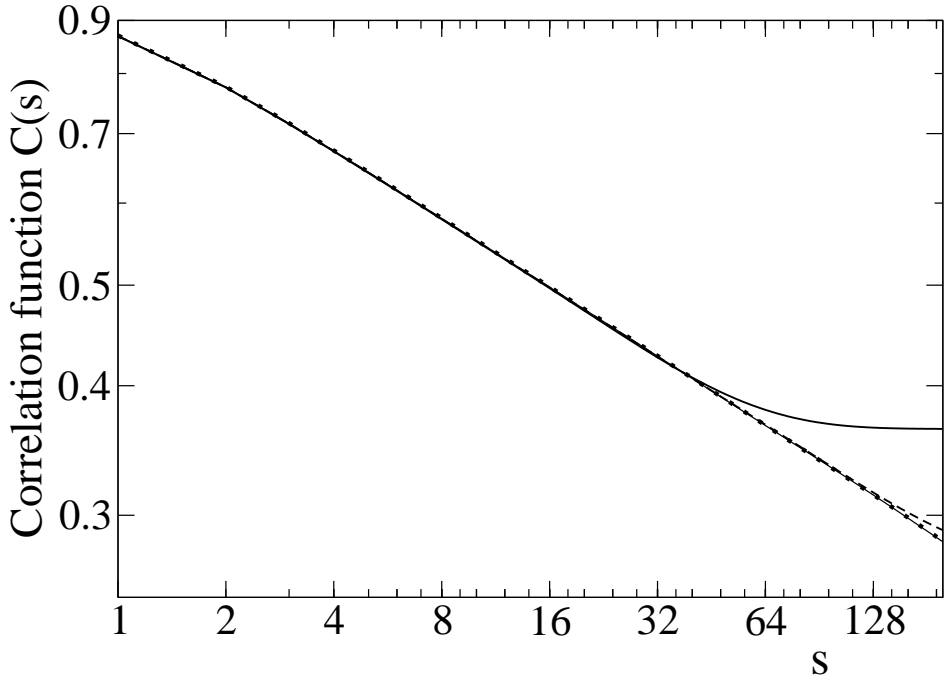}
\caption{Correlation function $C(s)$  in the superfluid phase regime,
$J=0.32$. Solid line $\chi=30$, dashed $\chi=100$ while the dots and the underlying
thin solid line correspond to   
 $\chi=150$ and $\chi=200$, respectively.
The last two sets almost coincide indicating the convergence with respect to $\chi$.
The linear behaviour in this log-log plot proves that the correlation function decays
algebraically. }
\label{corchi}
\end{figure}

It seems quite surprising that our results both for the energy 
and for the correlation 
function, Fig.~\ref{corchi}, converge 
well even in the superfluid regime where 
the gap in the excitation spectrum should vanish. We indeed observe such a convergence,
although apparently not fully exponential. We presently have no clear explanation of this behavior,
which is under study.
A plausible cause is  the enforcement of translational 
invariance at each step of the algorithm. That restricts the spanned Hilbert space
to the translationally invariant subspace killing, e.g.,
the possible excitations of low-frequency Bogolyubov modes in the superfluid regime.
We carefully checked that all results presented in this paper are converged with respect
to a modification of the (small) time step and an increase of $\chi.$

\begin{figure}[t]
\includegraphics[height=.5\textheight]{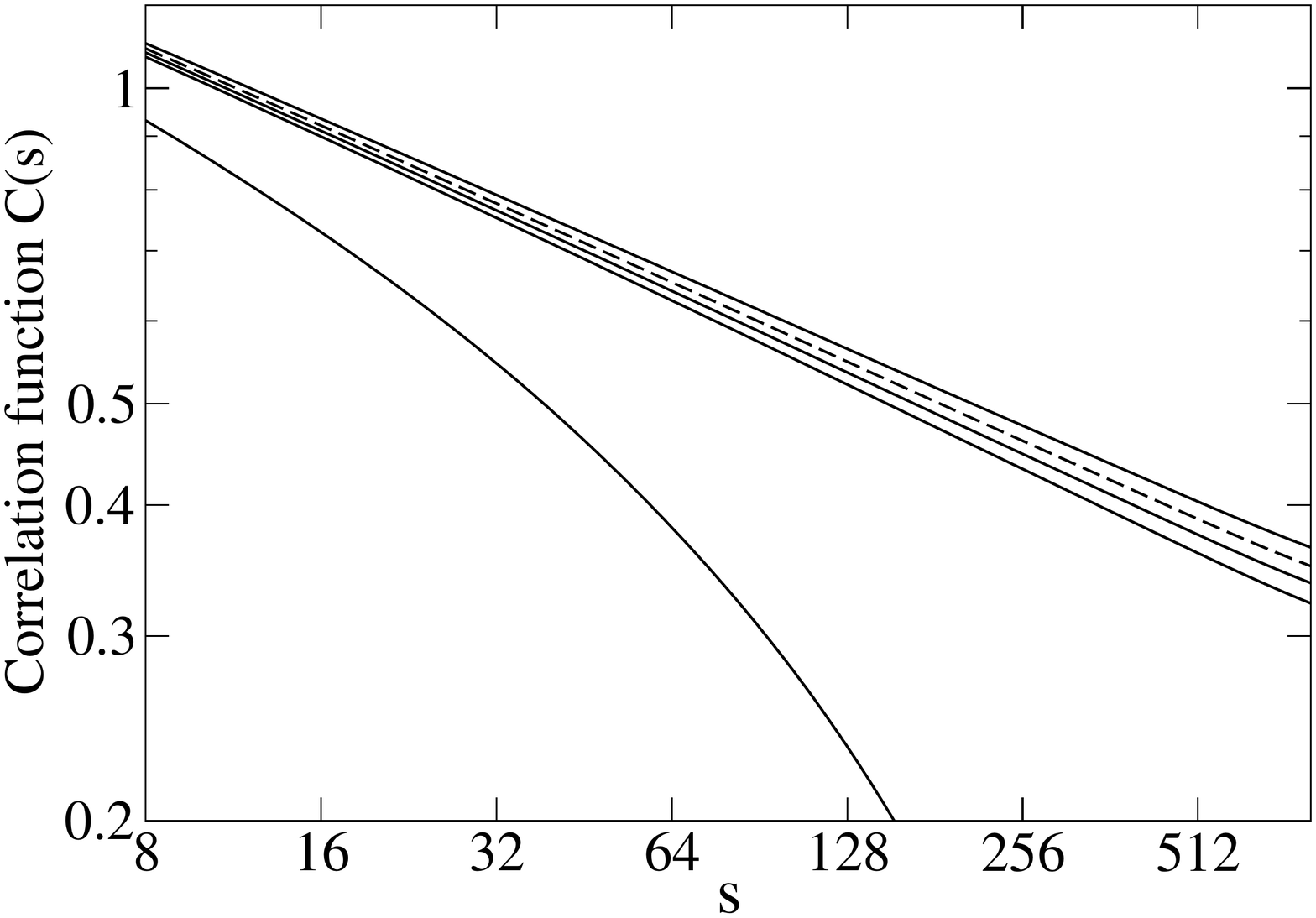}
\caption{Correlation functions $C(s)$ for different values of 
$J$ on a double logarithmic scale. From bottom to top: 
$J=0.26, 0.29, 0.293, 0.296, 0.30$. The $J=0.26$ curve in the deep Mott regime
regime corresponds to a Mott insulator. As expected it resembles an exponential. 
In the deep superfluid regime, the power law is very well obeyed, 
that allows for accurate fits. Close to the transition, even on the Mott side, $C(s)$
shows transient behaviour for small $s$, then a region of an approximate power law
behaviour, then for very large $s$ an exponential tail is expected in the Mott regime.
When approaching the critical point on the Mott side,
this exponential region takes place at larger and larger $s.$
In the range $s\leq 800$ shown in this plot, it is not visible
expect for the lowest $J=0.26$ curve.
The $J=0.296$ curve, the closest one to the critical curve $s^{-1/4},$ 
is shown by a dashed line. 
}
\label{correl}
\end{figure}

The convergence 
of the long-range correlation functions is of particular importance. 
As discussed
in \cite{km,kwm00}, at integer density, it should exhibit a 
power law $C(s)\propto s^{-K/2}$ for large $s$, on the superfluid side 
with $K$ increasing up to $K=0.5$ when the KT transition is approached. 
This is a behaviour typical 
for a Luttinger liquid characterising low energy excitations in 
the superfluid state. Fig.~\ref{correl} reveals that it is indeed the 
case. As expected, 
the power law behaviour no longer holds on the Mott side of the 
transition. Here, deep in the Mott regime, an exponential decay of correlations is observed.

\begin{table}
\begin{tabular}{ccc}
\hline
   \tablehead{1}{r}{b}{$s$}
  & \tablehead{1}{r}{b}{$J_c$ (DMRG) }
  & \tablehead{1}{r}{b}{$J_c$}   \\
\hline
$4\le s \le 8$ & $0.2874 \pm 0.0001$ &$0.2868 \pm 0.0006$\\
$8\le s \le 16$ &$0.2938 \pm 0.0001$ & $0.2933 \pm 0.0002$ \\
$16\le s \le 32$ &$0.2968 \pm 0.0003$ & $0.2955 \pm 0.0001$ \\
$32\le s \le 48$ & $0.3062 \pm 0.0003$ & $0.2970 \pm 0.0001$ \\
$48\le s \le 64$ & $0.3107 \pm 0.01\phantom{00}$ & $0.2975 \pm 0.0001$ \\
$100\le  s  \le 250$ & &$0.2980 \pm 0.0001$     \\  
$100\le  s  \le 500$ & &$0.2972 \pm 0.0001$ \\
$50\le  s  \le 500$ & &$0.2975 \pm 0.0001$ 
    \\  \hline
\end{tabular}
\caption{\label{tab} The location of the critical point $J_c,$ estimated from 
different ranges of $s$ value in the correlation function $C(s)$. 
The second column reproduces fits of \cite{kwm00}, the third column yields present results.}

\end{table}

To estimate the KT transition point, a special case must be paid to keep a constant unit
density. While for a finite system the particle number conservation can be explicitly used
to keep the density fixed this is not the case for the infinite version of the code \cite{vidali}.
Instead, a chemical potential, $\mu$, must be used and adjusted to keep a contant desided density.
Our simulations were performed for different values of $\mu$ and the correlation functions 
shown (and used later to extract the transition point) were obtained for $\rho$ well
within $[0.999,1.001]$ interval. To obtain slopes at $\rho=1$ an appropriate interpolation of
the results
has been used. We observed that the power law coefficient in that interval
changed weakly, the possible error in the transition point evaluation is included in the estimates
given below. 

The  correlation functions $C(s)$ obtained at different $J$ values are fitted
with a power law on the superfluid side of the transition point. 
Since in the Mott regime the behaviour of the correlation function hardly resembles 
a power law (see Fig.~\ref{correl}), we
make fits in different ranges of $s$ exactly following the procedure of  
K\"uehner et al~\cite{kwm00}. The obtained slopes 
(as a function of $J$) are interpolated to yield 
the transition point $J_c$. These points
are collected  in
Table I for different $s$ intervals. 
For comparison we reproduce also the results of 
 \cite{kwm00}. While earlier DMRG results seem to be quite sensitive to the 
 range of $s$ values, our fits  yield much closer and mutually consistent results. 
  The
 comparison presented shows the power and accuracy of the infinite TEBD algorithm.
 
Clearly, fits for small $s$ lead to misleading results, $C(s)$ for $s$ below few tens (say 30)
does not yet behave in the asymptotic power law way even in the superfluid regime.
Given the accuracy of the method we can extend the evaluation of $C(s)$ to much larger $s$ values,
yielding global fits reported at the bottom of Table I. Note that those fits are in decent
agreement with moderate $s$ values of the order of 50.

We cannnot, however, extend the calculation of $C(s)$ to much larger $s$ values of the order of
several thousands or more due to the numerical precision.

The KT transition point at unit density 
should be safely close to
the value given by the last fit $J_c=0.2975 \pm 0.0005$ where the (conservative) estimate
of the error is coming from the comparison with other fits. 
While the obtained value is quite close to that reported 
previously \cite{kwm00}, even our conservative error estimate makes the 
result more accurate by one order of magnitude.

Quite a similar approach can be used for a double occupancy case $\rho=2$. Here, using the
very same method,  we obtain a benchmark value $J_c(2)=0.175\pm0.002$. For that calculation
the possible maximal occupancy of each site had to be larger or equal to 6. We have checked by 
a comparison of selected runs with those for even larger occupancy that the assumed maximal
occupation is sufficient for the convergence.

\section{Conclusions}

To summarize, we have shown that the infinite TEBD algorithm \cite{vidali} 
may be efficiently applied to nontrivial challenging problems such as 
the determination of
 the KT critical point in the BHM. The accuracy 
 is  superior to that obtained by the standard DMRG approach.
Surprisingly, converged results are 
also obtained in the superfluid regime. The restriction of the algorithm to
the translationally invariant subspace, while justified for present ground state
studies, may restrict the application of the method \cite{vidali} to
 the real time evolution. There it could allow to 
treat ``exactly'' the dynamics of the quantum phase transition. 
Work in this direction is in progress.


\begin{theacknowledgments}
We acknowledge insightful discussions with Bogdan Damski and
Zbyszek Karkuszewski. We
are grateful for support by  Polish-French Polonium collaboration scheme as well 
as Marie Curie ToK  project COCOS (MTKD-CT-2004-517186).
Part of calculations were done in ICM UW under grant G29-10.  
J.Z. acknowledges support of Polish Government Research Grant for 2006-2009.
\end{theacknowledgments}


\end{document}